\documentclass[12pt]{article}

\usepackage[dvips]{graphicx}
\usepackage{latexsym}

\usepackage{amssymb,amsfonts,amsmath}
\usepackage{graphicx} 
\usepackage{indentfirst}

 \usepackage{bbm}

\parskip=\medskipamount

\newcommand {\cH}{{\cal H}}

\newcommand {\cK}{{\cal K}}
\newcommand {\cL}{{\cal L}}
\newcommand {\cM}{{\cal M}}
\newcommand {\cN}{{\cal N}}

\newcommand {\cP}{{\cal P}}

\def\a{\alpha}
\def \bi{\bibitem}

\def\b{\beta}
\def\c{\chi}
\def\d{\delta}
\def\e{\epsilon}
\def\f{\phi}
\def\g{\gamma}
\def\G{\Gamma}

\def\l{\lambda}

\def\o{\omega}
\def\p{\pi}
\def\q{\theta}
\def\r{\rho}
\def\s{\sigma}

\def\u{\upsilon}
\def\x{\xi}
\def\z{\zeta}

\def\F{\Phi}

\def\L{\Lambda}

\def\S{\Sigma}
\def\U{\Upsilon}

\def\rd{{\rm d}}
\def\ri{{\rm i}}

\newcommand{\ad}{{\dot{\alpha}}}                           
\newcommand{\bd}{{\dot{\beta}}}                            
\newcommand{\ve}{\varepsilon}                            

\newcommand{\pa}{\partial}                           
\newcommand{\hf}{\frac12}

\newcommand{\vf}{\varphi}
\newcommand{\sect}[1]{\setcounter{equation}{0}\section{#1}}

\newcommand{\be}{\begin{equation}}
\newcommand{\ee}{\end{equation}}
\newcommand{\bea}{\begin{eqnarray}}
\newcommand{\eea}{\end{eqnarray}}
\newcommand{\non}{\nonumber}
\newcommand{\1}{\underline{1}}
\newcommand{\2}{\underline{2}}

\def\dt#1{{\buildrel {\hbox{\LARGE .}} \over {#1}}}    

\newcommand{\bm}[1]{\mbox{\boldmath$#1$}}

\def\double #1{#1{\hbox{\kern-2pt $#1$}}}

\def\double #1{#1{\hbox{\kern-2pt $#1$}}}

\begin{document}

\begin{titlepage}

\begin{flushright}
October, 2007\\
\end{flushright}
\vspace{5mm}

\begin{center}
{\Large \bf  On superconformal projective hypermultiplets}
\end{center}

\begin{center}

{\large  
Sergei M. Kuzenko\footnote{{kuzenko@cyllene.uwa.edu.au}}
} \\
\vspace{5mm}

\footnotesize{
{\it School of Physics M013, The University of Western Australia\\
35 Stirling Highway, Crawley W.A. 6009, Australia}}  
~\\

\vspace{2mm}

\end{center}
\vspace{5mm}

\begin{abstract}
\baselineskip=14pt
Building on the five-dimensional constructions  in hep-th/0601177, 
we provide a unified description of  four-dimensional $\cN=2$ superconformal off-shell
multiplets in projective superspace, including a realization
in terms of $\cN=1$ superfields. In particular, superconformal polar multiplets 
are consistently defined for the first time.  We present new 4D $\cN=2$ superconformal 
sigma-models described by polar multiplets. Such sigma-models realize general 
superconformal couplings in projective superspace, but  involve an infinite tale 
of auxiliary $\cN=1$ superfields. The auxiliaries   should be eliminated by solving infinitely 
many algebraic nonlinear equations, and this is a nontrivial technical problem. 
We argue that the latter can be avoided by making use of supersymmetry considerations.  
All information about the resulting superconformal model 
(and hence  the associated superconformal cone)
is encoded in the so-called
canonical coordinate system for a K\"ahler metric, which was introduced by Bochner 
and Calabi in the late 1940s.
\end{abstract}
\vspace{1cm}

\vfill
\end{titlepage}

\newpage
\renewcommand{\thefootnote}{\arabic{footnote}}
\setcounter{footnote}{0}

\tableofcontents{}
\vspace{1cm}
\bigskip\hrule

\sect{Introduction}
Hyperk\"ahler manifolds  are known to be the target spaces for 
systems of 4D $\cN=2$ hypermultiplets in the case of rigid supersymmetry
\cite{A-GF}. In local supersymmetry, when the hypermultiplets
couple to $\cN=2$ supergravity, their target spaces have to be  quaternionic 
K\"ahler  \cite{BW}.
Unlike K\"ahler metrics, both hyperk\"ahler  and quaternionic 
K\"ahler metrics are difficult to construct explicitly. 
However, the results of \cite{A-GF,BW} 
imply that the existence of
regular (i.e. superspace)   techniques for formulating  supersymmetric nonlinear 
sigma-models with eight supercharges should be equivalent to a  formalism
to generate hyperk\"ahler  and quaternionic K\"ahler metrics.  This idea was one 
of the driving motivations  in the 1980s to look for 4D $\cN=2$ off-shell 
supersymmetric techniques, and the latter  quest has resulted in the creation of two powerful 
paradigms:\footnote{It was Rosly \cite{Rosly} who first realized, building on earlier ideas 
due to Witten \cite{Witten},  that the right superspace setting 
for 4D $\cN=2$ supersymmetric theories  is  {\it isotwistor superspace}
${\mathbb R}^{4|8} \times {\mathbb C}P^1 ={\mathbb R}^{4|8} \times S^2$
(following the terminology of \cite{RS}).
This superspace is called ``harmonic'' or ``projective,''
depending on the following two prerequisites:
(i) the supermultiplets selected to inhabit it; 
and (ii) the supersymmetric action principle chosen.}
harmonic superspace \cite{GIKOS,GIOS}  and projective superspace\footnote{See
\cite{GHR} for a related construction in two dimensions.}
\cite{KLR,LR1,LR2}.
The projective superspace approach is ideally suited for 
explicit construction of hyperk\"ahler metrics.
 
Remarkably, the problem of constructing arbitrary quaternionic
K\"ahler metrics is equivalent to that of generating
hyperk\"ahler metrics with special properties. 
As shown first by Swann \cite{Swann} (see also \cite{Galicki}),  there exists a one-to-one
correspondence between $4n$-dimensional quaternionic K\"ahler spaces
and $4(n+1)$-dimensional hyperk\"ahler manifolds possessing a homothetic 
Killing vector (implying the fact that the
isometry group includes a subgroup SU(2) that rotates  
the three complex structures). In the physics literature, such hyperk\"ahler spaces
are known as ``hyperk\"ahler cones'' \cite{deWRV}, and they turn out to be  the target spaces
for 4D $\cN=2$ superconformal sigma-models (see \cite{deWKV,deWRV,RVV} 
and references therein). Given a $4(n+1)$-dimensional hyperk\"ahler cone, 
the corresponding $4n$-dimensional quaternionic K\"ahler space is obtained
by implementing the Swann reduction \cite{Swann,Galicki}. 
At the sigma-model level, this was elaborated in detail in \cite{deWRV}.\footnote{For 
the construction of quaternionic K\"ahler metrics from harmonic superspace, 
see \cite{GIOS,IV} and references therein.} 

Thus, to generate arbitrary hyperk\"ahler cones, it is sufficient to construct 
all possible superconformal sigma-models described in terms of  various off-shell realizations
of the 4D $\cN=2$ massless scalar multiplet. So far, this has thoroughly been elaborated 
\cite{deWRV,RVV} for only the simplest off-shell realization -- $\cN=2$ tensor multiplet
\cite{Wess,N=2tensor,BS}
(see \cite{deWS} for a detailed study of tensor multiplets in $\cN=2$ supergravity).
General couplings for $\cN=2$ tensor multiplets were actually given in the
foundational  work on projective superspace \cite{KLR}, 
and  even earlier in \cite{LR83}.
As is known, the use of tensor multiplets allows one to generate very restrictive couplings.
At the same time, the most interesting multiplet in projective superspace 
is the so-called polar multiplet \cite{LR1,LR2,G-RLRvUW}, for it is believed to allow the 
most general sigma-model couplings\footnote{The polar multiplet is the
projective-superspace analogue of the $q^+$-hypermultiplet in harmonic superspace 
\cite{GIKOS}, see \cite{K98} for a detailed discussion of the relationship 
between these two approaches.}  \cite{LR1,LR2}.
The superconformal description of polar multiplets, as well as general 
superconformal couplings for polar multiplets, have been given only recently 
in the context of five-dimensional $\cN=1$ supersymmetry \cite{K}. 
The present paper is aimed, in part,  at extending the results of \cite{K} to four 
dimensions.\footnote{Such an extension is very natural. But since the 5D 
superspace notation and the  corresponding superconformal algebar F(4) \cite{Kac},
which were use in \cite{K}, are somewhat exotic, 
the 4D $\cN=2$ implications of the results in \cite{K}
do not seem to be transparent even for some experts.}

The main thrust of this paper is actually to address the following technical issue. 
When realized in terms of $\cN=1$ superfields, the polar multiplets
involve an infinite tale of auxiliary unconstrained superfields, along 
with two physical superfields. In nonlinear sigma-models, 
elimination of the auxiliaries requires solving an infinite set of algebraic 
nonlinear equations, and this is hard. We are going to demonstrate that 
this nontrivial  problem can completely be avoided by making use of powerful
supersymmetry considerations.  Conceptually, this will be similar to 
the recent analysis given for the $\cN=2$ supersymmetirc sigma-models 
on tangent bundles on Hermitian symmetric spaces \cite{AKL2}. 

This paper is organized as follows. 
In section 2, we start by recalling the 4D $\cN=2$ superconformal 
kinematics, and then continue on to define superconformal projective 
multiplets and give several important examples. 
The superconformal action principle is also discussed.
Superconformal sigma-models  are presented in section 3. 
As an illustration, here we review the models for tensor  
(and, more generally, $O(2n)$) multiplets, 
which have already appeared in the literature.  
A new family of superconformal sigma-models for polar hypermultiplets 
is introduced. The latter theories provide general superconformal 
sigma-model couplings in projective superspace. 
In section 4, we discuss the reduction of $\cN=2$ superconformal 
multiplets to $\cN=1$ superfields. In section 5, we consider a large class 
of polar hypermultiplet theories, which include the superconformal sigma-models 
as a subclass, and address the problem of eliminating the auxiliary degrees 
of freedom. The specific features of the superconformal sigma-models are  
analyzed in section 6.  Finally, some facts about $\cN$-extended superconformal 
Killing vectors are collected in the appendix. 
  
\sect{4D \mbox{$\bm{ \cN=2}$} superconformal formalism}
In this section, we introduce various superconformal projective 
multiplets and discuss  the  manifestly superconformal action principle. 
We start by recalling the key points of the superconformal 
formalism in 4D $\cN=2$ superspace 
 ${\mathbb R}^{4|8}$ parametrized  
by  coordinates  $ z^A = (x^a,  \q^\a_i, {\bar \q}^i_\dt{\a} )$, 
where $i=\1, \2 $. 

\subsection{Superconformal Killing vectors}
Here we build on the formalism developed in \cite{KT} (see also \cite{Park}).
By definition, a superconformal  Killing vector\footnote{The concept of superconformal Killing  
vectors \cite{Sohnius2,Lang,BPT,Shizuya,BK,HH,West}
is extremely useful for various studies of 
superconformal theories in four, five and six 
dimensions, see e.g.  \cite{Osborn,Park,KT,K}.}
\be
\x = {\overline \x} =\x^A (z)D_A
= \x^a (z) \,\pa_a + \x^\a_i (z)\,D^i_\a
+ {\bar \x}_{\dt \a}^i (z)\, {\bar D}^{\dt \a}_i
\ee   
obeys the condition 
\bea
{\bar D}_{\dt \a}^i \F =0 \quad \longrightarrow \quad 
{\bar D}_{\dt \a}^i (\x \,\F )=0~, 
\eea
for an arbitrary chiral superfield $\F$.
This condition implies the fulfillment of eq. (\ref{4Dmaster}) and also 
\be
[\x \;,\; D^i_\a ] = - (D^i_\a \x^\b_j) D^j_\b
= { \o}_\a{}^\b  D^i_\b - 
\bar{ {\s}} \, D^i_\a
- {\L}_j{}^i \; D^j_\a ~.
\label{4DmasterN=2} 
\ee
The latter relation corresponds to the choice $\cN=2$ in  eq. (\ref{4Dmaster2}).
The parameters of ($z$-dependent) Lorentz ${\o}$ 
and scale--chiral ${\s}$ transformations are
\be
{\o}_{\a \b}(z) = -\frac{1}{2}\;D^i_{(\a} \x_{\b)i}\;,
\qquad {\s} (z) = \frac{1}{4}
{\bar D}^{\dt \a}_i {\bar \x}_{\dt \a}^{ i} 
\label{lor,weylN=2}
\ee
and they can be seen  to be chiral
\be
{\bar D}^{\dt \a}_{ i} {\o}_{\a \b}~=~ 0\;,
\qquad {\bar D}^{\dt \a}_{ i} {\s} ~=~0\;.
\ee
The parameters ${\L}_j{}^i$ defined by
\bea
{\L}_j{}^i (z) &=& \hf \Big(D^i_\a \x^\a_j - \hf \d^i_j D^k_\a \x^\a_k \Big)
=-\hf \Big({\bar D}^{\dt \a}_j {\bar \x}^i_{\dt \a} 
-\hf \d^i_j {\bar D}^{\dt \a}_k {\bar \x}^k_{\dt \a} \Big)
~,\non \\
&&\qquad 
\L^{ij}=\L^{ji}~,
\qquad \overline{\L^{ij} } = \L_{ij}  
\label{lambdaN=2}
\eea
correspond to   SU(2 ) transformations.
One can readily check the identity 
\be
D^k_\a {\L}_j{}^i = -2 \Big( \d^k_j D^i_\a 
-\frac{1}{2} \d^i_j D^k_\a  \Big) {\s}~,
\label{an1N=2}
\ee
and therefore 
\bea
D^{(i}_\a \L^{jk)} = {\bar D}^{(i}_{\dt \a} \L^{jk)} =0~.
\label{L-an}
\eea

A primary superfield $\cH (z)$ (with its Lorentz and SU(2) indices suppressed) 
is defined to possess the superconformal transformation 
\bea
\d \cH &=& - \Big( \x 
+{\o}_{\a}{}^{ \b}  M_{\b}{}^{ \a}
+ \bar{\o}_{\dt \a}{}^{  \dt \b}  
\bar{M}_{\dt \b}{}^{ \dt \a} 
+ {\L}_i{}^j 
R_j{}^i  
 + 2\big( p\, \s + q\, \bar{\s} )
\Big)
\cH~.
\eea
Here $M_{\a}{}^{ \b}$ and $\bar{M}_{\dt \a}{}^{ \dt \b}$ are 
the Lorentz generators,  and $R^i{}_j$  the generators of SU(2).
The parameters $p$ and $q$ determine the dimension  $(p+q)$ of the superfield
and its ${\rm U}(1)_R$ charge proportional to $(p -q)$.

${}$Following  \cite{Rosly,GIKOS,KLR},
it is robust to make use of  
an isotwistor $u^{+i} \in {\mathbb C}^2 \setminus \{0\}$
that allows one to introduce  a subset of strictly anti-commuting  spinor covariant derivatives,
in accrodance with (\ref{spinor-c-d}), 
\bea
D^+_\a =D^i_\a \,u^+_i ~, \quad {\bar D}^+_{\dt \a} ={\bar D}^i_{\dt \a} \,u^+_i~ 
\qquad \{ D^+_\a , D^+_\b\}=\{ {\bar D}^+_{\dt \a} , {\bar D}^+_{\dt \b} \}
=\{ D^+_\a , {\bar D}^+_{\dt \b}\}=0~.~~~
\eea
Hence, one can define so-called {\it analytic} superfields\footnote{Such superfields 
were called {\it isochiral} in \cite{RS}.} 
constrained by 
$D^+_\a Q = {\bar D}^+_{\dt \a} Q=0$. 

Let us introduce 
\be
\L^{++} =\L^{ij} \,u^+_i u^+_j~.
\label{L++}
\ee 
It follows from (\ref{L-an}) that $\L^{++} $ is analytic, 
\be
D^+_\a \L^{++}  = {\bar D}^+_{\dt \a} \L^{++} =0~.
\ee

In addition to $u^+_i$, it is also useful to introduce an auxiliary 
isotwistor $u^-_i$ obeying the only condition
\be
(u^+u^-) = u^{+i}u^-_i \neq 0~.
\label{u-}
\ee
Of course, with  $u^-_i$ fixed, this condition  is satisfied 
only on an open subset of the isotwistor space ${\mathbb C}^2 \setminus \{0\}$.
With its aid, we introduce the  isotwistor derivatives
(compare with  \cite{GIKOS})
\bea
D^{++}=u^{+i}\frac{\partial}{\partial u^{- i}} ~,\qquad
D^{--}=u^{- i}\frac{\partial}{\partial u^{+ i}} ~,
\label{5}
\eea
and the spinor covariant derivatives 
\be
D^-_\a =[D^{--}, D^+_\a]~, \qquad
{\bar D}^-_{\dt \a} =[D^{--}, {\bar D}^+_{\dt \a}]~.
\ee
Since $u^+_i$ and $u^-_i$ form a linearly independent basis for ${\mathbb C}^2$, 
the superconformal Killing vector can also be represented as follows:
\bea
\x = {\overline \x} = \x^a (z) \,\pa_a 
-\frac{1}{(u^+u^-)} \Big( \x^{+\a} D^-_\a + {\bar \x}^{+\dt \a} {\bar D}^-_{\dt \a} \Big)
+\frac{1}{(u^+u^-)} \Big( \x^{-\a} D^+_\a + {\bar \x}^{-\dt \a} {\bar D}^+_{\dt \a} \Big)~,~~~
\label{xi-2}
\eea
with $\x^{\pm \a} =\x^{\a i} \,u^+_i $ and 
${\bar \x}^{+ \dt \a} ={\bar \x}^{\dt \a  i} \,u^+_i$.

Using eq. (\ref{an1N=2}) one can show that 
the following combination
\be
\S =  \frac{\L^{+-}}{(u^+u^-)} \,  + {\s} 
+\bar{ {\s} }~, \qquad 
\L^{+-} = \L^{ij} \, u^+_i u^-_j 
\label{Sigma0}
\ee
possesses the properties
\be
D^+_\a \, \S = {\bar D}^+_{\dt \a} \, \S=0~, \qquad 
D^{++} \S =  \frac{\L^{++}}{(u^+u^-)}~,
\label{Sigma}
\ee
and thus $\S$ is analytic.\footnote{There are natural analogs of   $\L^{++}$ and $\S$ 
in the harmonic-superspace approach \cite{GIOS-conf,GIOS}.}
Now, the (supervolume-preservation)  identity (see, e.g. \cite{BKT}) 
\be
(-1)^A D_A\, \x^A =0
\label{volume1}
\ee
can be rewritten in the form 
\bea
\pa_a  \x^a
+\frac{1}{(u^+u^-)} \Big(D^-_\a  \x^{+\a}  + {\bar D}^-_{\dt \a}{\bar \x}^{+\dt \a} 
-D^{--}\L^{++} \Big) = 2\S~.
\label{volume2}
\eea

Eq. (\ref{4DmasterN=2}) implies 
\be
\d D^+_\a \equiv \Big[ \x - \frac{\L^{++}}{(u^+u^-)}
D^{--}  ,   D^+_{ \a} \Big]
=  {\o}_{ \a}{}^{ \b}\, D_{ \b}^+
- \Big( {\s}  +   \frac{\L^{+-}}{(u^+u^-)} \, D^+_{ \a}\Big)~,
\label{master-+}
\ee
and similarly for $\d {\bar D}^+_{\dt \a} $.

\subsection{Superconformal projective multiplets: Definition}
In defining 4D $\cN=2$ superconformal multiplets in projective superspace,
we closely follow the formulation of 5D superconformal off-shell multiplets
given in \cite{K},
and the  subsequent extension  
for 5D $\cN=1$ AdS superspace \cite{KT-M}.  

A superconformal projective multiplet of weight $n$,
$Q^{(n)}(z,u^+)$, is a superfield that 
lives on  ${\mathbb R}^{4|8}$, 
is holomorphic with respect to 
the isotwistor variables $u^{+}_i $ on an open domain of 
${\mathbb C}^2 \setminus  \{0\}$, 
and is characterized by the following conditions:\\
(i) it obeys the analyticity constraints 
\be
D^+_{\a} Q^{(n)} ={\bar D}^+_{\dt \a} Q^{(n)} =0;
\label{ana}
\ee  
(ii) it is  a homogeneous function of $u^+$ 
of degree $n$, that is  
\be
Q^{(n)}(z,c\,u^+)\,=\,c^n\,Q^{(n)}(z,u^+)~, \qquad c\in \mathbb{C}^*~;
\label{weight}
\ee
(iii) it possesses the  superconformal transformation law:
\be
\d Q^{(n)} = - \Big(  \x -  \frac{\L^{++}}{(u^+u^-)} D^{--} \Big) \, Q^{(n)} 
-n \,\S \, Q^{(n)} ~.
\label{harmult1}
\ee
As a consequence of eqs. (\ref{Sigma}) and (\ref{master-+}), 
the variation $\d Q^{(n)} $ is analytic.
By construction, $Q^{(n)}$ is independent of the auxiliary isotwistor $u^-_i$,
\be
\frac{\pa}{\pa u^{-i}}\, Q^{(n)}=0 \quad \longrightarrow \quad
D^{++}Q^{(n)}=0~.
\ee 
Eq. (\ref{weight}) implies that $\d Q^{(n)}$ 
is also independent of $u^-$,
\bea
\frac{\pa}{\pa u^{-i}} \,
\d Q^{(n)} =0~, 
\eea
although separate contributions to the right-hand side  of (\ref{harmult1})
involve $u^-$. In order for eq. (\ref{weight}) 
(and also eq. (\ref{tilde-conj})) to be unambiguous, in what follows we restrict 
the weight $n$ to be integer.

Using the natural projection $\p\!\!: {\mathbb C}^2 \setminus  \{0\} \to  {\mathbb C}P^1$, 
the superconformal projective multiplets can be reformulated as tensor fields
that live in ${\mathbb R}^{4|8} \times {\mathbb C}P^1$ and are holomorphic 
on an open domain of ${\mathbb C}P^1$, see below.
In the harmonic-superspace approach \cite{GIKOS,GIOS}, 
one has to deal with smooth tensor fields on  ${\mathbb R}^{4|8} \times S^2$ 
which are globally defined on $S^2$.
The projective-superspace action \cite{KLR,S}
does not require the Lagrangian (and, hence,  the matter superfields appearing in the Lagrangian) 
to be globally defined over ${\mathbb C}P^1$. In practice this often gives some more freedom, 
say, for sigma-model building.

Simplest superconformal projective multiplets are homogeneous polynomials in $u^+$
\be
H^{(n)}(z,u)\,=\,u^+_{i_1}\cdots u^+_{i_n}\,H^{i_1\cdots i_n}(z)~.
\label{O(n)2}
\ee
${}$Following the terminology of \cite{G-RLRvUW}, 
they will be called  $O(n)$ multiplets.
Such multiplets are globally defined on ${\mathbb C}^2 \setminus  \{0\}$.
The analyticity constraints (\ref{ana}) are equivalent to 
\bea
D^{(j}_\a H^{i_1\cdots i_n)} = {\bar D}^{(j}_{\dt \a} H^{i_1\cdots i_n)} = 0~.
\label{ana2}
\eea
The transformation law (\ref{harmult1}) is equivalent to 
\bea
\d H^{i_1\cdots i_n} =- \x H^{i_1\cdots i_n}
-\sum_{k=1}^{n} \L_j{}^{i_k} H^{j i_1\cdots  \widehat{i_k} \cdots i_n} 
-n(\s +{\bar \s}) H^{i_1\cdots i_n}~,
\eea
where the notation $\widehat{i_k} $ means that the corresponding index is missing.
The latter transformation law is uniquely determined by the constraints
(\ref{ana2}).
It should be pointed out that the case $n=1$ corresponds 
to an on-shell Fayet-Sohnius hypermultiplet
\cite{Sohnius}, $n=2$ to an off-shell tensor multiplet \cite{N=2tensor}.
In the super-Poincar\'e case, 
general  $O(n)$ multiplets, with  $n>2$,  were studied in \cite{KLT,GIO,LR2}.

The complex conjugate of an analytic  superfield $Q^{(n)}$ is not analytic.
However, one can introduce a generalized,  analyticity-preserving 
conjugation  \cite{Rosly,GIKOS,KLR}, 
$Q^{(n)} \to \widetilde{Q}^{(n)}$, defined as (see also \cite{KT-M})
\be
\widetilde{Q}^{(n)} (u^+)\equiv \bar{Q}^{(n)}\big(\widetilde{u}^+\big)~, 
\qquad \widetilde{u}^+ = {\rm i}\, \s_2\, u^+~, 
\ee
with $\bar{Q}^{(n)}$ the complex conjugate of $Q^{(n)}$.
Its fundamental property is
\bea
\widetilde{ {D^+_{\a} Q^{(n)}} }=-(-1)^{\e(Q^{(n)})}\, {\bar D}^{+}_{\dt \a}
 \widetilde{Q}{}^{(n)}~, \qquad 
 \widetilde{ {\bar D}^{+}_{\dt \a} Q^{(n)} } =  (-1)^{\e(Q^{(n)})}\, D^{+}_{ \a}
 \widetilde{Q}{}^{(n)}~.  
\eea
One can show
\be
\widetilde{\widetilde{Q}}{}^{(n)}=(-1)^nQ^{(n)}~,
\label{tilde-conj}
\ee
and therefore real supermultiplets can be consistently defined when 
$n$ is even.
In what follows, $\widetilde{Q}^{(n)}$ will be called the smile-conjugate of 
${Q}^{(n)}$.

By smile-conjugating the transformation law (\ref{harmult1}), one can see that 
$\widetilde{Q}^{(n)}$ is  a superconformal projective multiplet of weight $n$.

\subsection{Superconformal projective multiplets: Examples}
Consider the natural projection
$\p\!\!: {\mathbb C}^2 \setminus  \{0\} \to  {\mathbb C}P^1$. 
The isotwistor variables $u^+_i$ provide homogeneous global coordinates 
for  points in ${\mathbb C}P^1$. Thus, any analytic superfield corresponds 
to a supermultiplet living in ${\mathbb R}^{4|8} \times {\mathbb C}P^1$.
Instead of $u^+_i$, it is often useful to deal with an inhomogeneous complex 
coordinate $\z$ which is defined  locally and is invariant under projective rescalings
$u^+_i \to c\, u^+_i $, with $ c\in \mathbb{C}^*$.
Then, one should replace $Q^{(n)}(z,u^+)$ with a new superfield 
$Q^{[n]}(z,\z) \propto Q^{(n)}(z,u^+)$, where $Q^{[n]}(z,\z) $ is  holomorphic 
with respect to  $\z$. 
As is demonstrated below, the precise definition of $Q^{[n]}(z,\z) $ depends on 
the projective supermultiplet under consideration.  
 It is standard to cover
 ${\mathbb C}^2 \setminus  \{0\} $ 
 by two open charts:
(i) the north chart characterized by $u^{+\1}\neq 0$;
(ii) the south chart with  $u^{+\2}\neq 0$.
In discussing various supermultiplets, 
our consideration below will be restricted to the north chart.

Since  $u^{+\1}\neq 0$ in the north chart, 
it is natural to introduce 
a projective-invariant complex variable $\z \in \mathbb C$
as follows:
\bea
u^{+i} =u^{+\1}(1,\z) =u^{+\1}\z^i ~,\qquad 
\z^i=(1,\z)~, \qquad \z_i= \ve_{ij} \,\z^j=(-\z,1)~.
\label{zeta}
\eea
Any projective multiplet $Q^{(n)}$ and its
superconformal variation (\ref{harmult1}) do not depend on $u^-$, 
and thus we can make a convenient choice for the later.
It is useful to choose
\be
u^-_i =(1,0) ~, \qquad   \quad ~u^{-i}=\ve^{ij }\,u^-_j=(0,-1)~.
\label{fix-u-}
\ee   
${}$For the analytic transformation parameters $\L^{++}$ 
(\ref{L++}) and $\Sigma$ (\ref{Sigma0}), we then have
\bea
\L^{++} =\big(u^{+\1}\big)^2 {\L}^{++} (\z)~, \quad 
{\L}^{++} (\z)&=& {\L}^{ij} \,\z_i \z_j
=  {\L}^{\1 \1 }\, \z^2 -2  {\L}^{\1 \2}\, \z 
+ {\L}^{\2 \2} ~,\non \\
\S=\S (\z) ~, \quad \qquad \qquad \quad 
\S(\z)&=& {\L}^{\1 i} \,\z_i + {\s} +{\bar \s}
=  - {\L}^{\1 \1} \,\z + {\L}^{\1 \2} + {\s} +\bar \s ~.~~
\label{L++Sigma}
\eea

An arctic multiplet\footnote{We use the terminology introduced in 
\cite{G-RLRvUW} for various projective multiplets
in the super-Poincar\'e case.}
 of weight $n $ is defined to be holomorphic 
on the north chart. It can be represented as 
\bea
\U^{(n)} (z, u) =  (u^{+\1})^n\, \U^{[n]} (z, \z) ~, \quad \qquad
\U^{ [n] } (z, \z) = \sum_{k=0}^{\infty} \U_k (z) \z^k~.
\label{arctic1}
\eea
The superconformal transformation law of $\U^{[n]}$ can be derived from 
eq. (\ref{harmult1}) to be
\bea
\d \U^{[n]} = - \Big(  \x &+&  {\L}^{++} (\z)\,\pa_\z \Big) \U^{[n]}
- n\,\S (\z) \, \U^{[n]}~.
\label{arctic2}
\eea
This transformation law is analogous to that  given in \cite{K} in five dimensions. 

The smile-conjugate of $ \U^{(n)}$ is said to be 
an antarctic multiplet of weight $n $. It proves to be  holomorphic on the south
chart, while  in the north chart it has the form
\bea
\widetilde{\U}^{(n)} (z, u) &=&  (u^{+\2})^n\, \widetilde{\U}^{[n]} (z, \z)
=  (u^{+\1})^n \z^n\, \widetilde{\U}^{[n]} (z, \z)~,  \non \\
\widetilde{\U}^{[n]} (z, \z) &=& \sum_{k=0}^{\infty} (-1)^k {\bar \U}_k (z)
\frac{1}{\z^k}~,
\label{antarctic1}
\eea
with $ {\bar \U}_k$ the complex conjugate of $U_k$.
In accordance with  (\ref{harmult1}), 
its superconformal transformation is as follows (compare with the 5D case \cite{K}):
\bea
 \d \widetilde{\U}^{[n]} =  
- \frac{1}{\z^n}\Big(  \x &+&  {\L}^{++} (\z) \,\pa_\z \Big) (\z^n\,\widetilde{\U}^{(n)} )
-n\,\S (\z) \,\widetilde{\U}^{(n)} ~.
\label{antarctic2}
\eea
The arctic multiplet $ \U^{[n]} $ and its smile-conjugate $\widetilde{\U}^{(n)} $ 
constitute a polar multiplet.

In the case of a real $O(2n)$ multiplet, it can be represented as 
\bea
H^{(2n)} (z,u^+) &=& 
\big({\rm i}\, u^{+\1} u^{+\2}\big)^n H^{[2n]}(z,\z) =
\big(u^{+\1}\big)^{2n} \big({\rm i}\, \z\big)^n H^{[2n]}(z,\z)~, 
\non \\
H^{[2n]}(z,\z) &=&
\sum_{k=-n}^{n} H_k (z) \z^k~,
\qquad  {\bar H}_k = (-1)^k H_{-k} ~. 
\label{o2n1}
\eea
In accordance with  (\ref{harmult1}), 
the  superconformal transformation  of $H^{[2n]} $ is
\bea
 \d  H^{[2n]} &=&  
- \frac{1}{\z^n}\Big(  \x +  {\L}^{++} (\z) \,\pa_\z \Big) (\z^n H^{[2n]} )
-2n \,\S (\z)\, H^{[2n]} ~,
\label{o2n2}
\eea
analogous to the five-dimensional transformation law \cite{K}.
In a similar way one can introduce complex $O(2n+1)$ 
multiplets.

${}$Finally, let us consider a real tropical multiplet of weight $2n$.
\bea
U^{(2n)} (z,u^+) &=& 
\big({\rm i}\, u^{+\1} u^{+\2}\big)^n U^{[2n]}(z,\z) =
\big(u^{+\1}\big)^{2n} \big({\rm i}\, \z\big)^n U^{[2n]}(z,\z)~, \non \\
U^{[2n]}(z,\z) &=&
\sum_{k=-\infty}^{\infty} U_k (z) \z^k~,
\qquad  {\bar U}_k = (-1)^k U_{-k} ~. 
\label{trop-nj}
\eea
Its superconformal transformation  copies (\ref{o2n2}).
The case $n=1$ corresponds to supersymmetric Lagrangians, 
see below. A tropical multiplet with $n=0$ is used to describe 
the prepotential for a massless vector multiplet.

In terms of the superfield $Q^{[n]}(z,\z) $, the analyticity condition  
(\ref{ana}) takes the form
\be
D^{\2}_{\a}Q^{[n]}(\z)=\z\,D^{\1}_{\a}Q^{[n]}(\z)~, \qquad 
{\bar D}_{\2}^{\ad}Q^{[n]}(\z)=-\frac{1}{\z}\,{\bar D}_{\1}^{\ad}Q^{[n]}(\z)~.
\label{ancon}
\ee
This relation implies  that the dependence
of the component superfields 
$Q_k $ of $Q^{[n]}(\z)$,
\bea
Q^{[n]}(z,\z) &=&
\sum_{k=-\infty}^{\infty} Q_k (z) \z^k~,
\label{series}
\eea
on $\q^\a_{\2}$ and ${\bar \q}^{\2}_{\dt \a}$ 
is uniquely determined in terms 
of their dependence on $\q^\a_{\1}\equiv \q^\a$
and ${\bar \q}^{\1}_{\dt \a}\equiv {\bar \q}_{\dt \a}$.  In other words, 
the projective superfields depend effectively 
on half the Grassmann variables which can be choosen
to be the spinor coordinates of 4D, $\cN=1$ superspace.
If the series (\ref{series}) terminates from below, 
then the two lowest components are constrained 
$\cN=1$ superfields. In particular, in the case of the arctic multiplet
(\ref{arctic1}), $\F:=\U_0|$ is chiral, ${\bar D}_{\dt \a} \F=0$, 
and $\S:=\U_1|$ is complex linear, ${\bar D}^2 \S=0$.

\subsection{Superconformal action }
Let $L^{++}(z,u^+) \equiv L^{(2)}(z,u^+)$ be 
a real superconformal projective multiplet of weight two. Following \cite{K},
we are going to  demonstrate that the 
action functional\footnote{In the super-Poincar\'e case,
this action was introduced in \cite{KLR}.
It was re-formulated in a 
manifestly  projective-invariant form in \cite{S}.} 
\bea
S= \frac{1}{ 2\pi}
\oint 
\frac{u_i^+\rd u^{+i}}{ (u^+u^-)^4}\int\rd^4 x\,({D}^-)^4L^{++}(z,u^+)
\Big| \!\Big| ~, 
\qquad (D^-)^4 =\frac{1}{16} (D^-)^2 ({\bar D}^-)^2 ~~
\label{Action}
\eea
is invariant under arbitrary  superconformal transformations.
Here the line integral  is carried out over a closed contour,
$\g =\{u^+_i(t)\}$,
in the space of $u^+$ variables. 
The integrand in (\ref{Action}) involves a constant (i.e. $t$-independent)
isotwistor $u^-_i$ subject to the only condition that $u^+(t)$ and $u^-$ form 
a linearly independent basis at each point of the contour $\g$, that is, 
eq. (\ref{u-}) holds at each point of the contour. 

In  (\ref{Action}), the double-bar notation,  $U{\double |}$, denotes the $\q$-independent
component of a $\cN=2$ superfield $U(x,\q_i, {\bar \q}^i)$.
Below, we will also a single-bar notation, $U|$, to denote the $\cN=1$ 
projection of $U$. Thus 
\be
U{\double |} = U(x,\q_i, {\bar \q}^i)\Big|_{\q_i={\bar \q}^i=0}~, \qquad 
U| = U(x,\q_i, {\bar \q}^i)\Big|_{\q_{\2}={\bar \q}^{\2}=0}~.
\ee

Action (\ref{Action}) is invariant under arbitrary
projective transformations of the form
\be
(u_i{}^-\,,\,u_i{}^+)~\to~(u_i{}^-\,,\, u_i{}^+ )\,R~,~~~~~~R\,=\,
\left(\begin{array}{cc}a~&0\\ b~&c~\end{array}\right)\,\in\,{\rm GL(2,\mathbb{C})}~.
\label{projectiveGaugeVar}
\ee
This gauge-like  symmetry implies that the action is actually independent of $u^-_i$. 
Using the representation (\ref{xi-2}) along with the analyticity conditions, 
the transformation law (\ref{harmult1}) with $n=2$ gives
\be
\d L^{++} = - \Big\{ 
\x^a \pa_a
-\frac{1}{(u^+u^-)} \Big( \x^{+\a} D^-_\a + {\bar \x}^{+\dt \a} {\bar D}^-_{\dt \a} 
+\L^{++} D^{--} \Big) \Big\} L^{++} 
-2 \,\S \, L^{++} ~.
\label{L++1}
\ee
Making here use of eq. (\ref{volume2})  leads to 
\bea
\d L^{++} = -\pa_a \big(\x^a L^{++}\big)
&-&\frac{1}{(u^+u^-)} \Big\{ D^-_\a\big( \x^{+\a}L^{++} \big)  + 
{\bar D}^-_{\dt \a}  \big( {\bar \x}^{+\dt \a} L^{++}\big)\Big\} \non \\
&+&\frac{1}{(u^+u^-)}D^{--}\big(\L^{++} L^{++}\big)~.
\eea
It remains to note the idenity (see \cite{KT-M} for a related discussion)
\bea
\frac{(\dt{u}^+u^+)}{(u^+u^-)^5}\,
D^{--}\big(\L^{++} L^{++}\big)&=&
-\frac{\rm d}{{\rm d}t} \Big( \frac{ \L^{++}L^{++} }{(u^+u^-)^4} \Big)~,
\eea
where $(\dt{u}^+u^+)\, {\rm d}t = u_i^+\rd u^{+i}$ is part of the line integral measure 
in (\ref{Action}). Since the line integral 
in (\ref{Action}) corresponds to a closed contour, the action is  seen to be invariant. 

We can now formulate a general superconformal Lagrangian:
\be 
L^{++} (z,u^+) = L \Big(Q^{(n)}(z,u^+) \Big)~, \qquad
L \Big(Q^{(n)}(z,c\,u^+) \Big) =c^2\, L \Big(Q^{(n)}(z,u^+) \Big)~.
\label{Lagrangian-regular}
\ee
Here the dynamical variables $Q^{(n)}$ are 
superconformal projective multiplets.

\subsection{Projective gauge fixing}
Without loss of generality, one can assume that the integration contour in 
(\ref{Action}) does not pass through the ``north pole'' $u^{+i} \sim (0,1)$.
Then, one can introduce the complex variable $\z$ as in (\ref{zeta}), 
and fix the projective invariance (\ref{projectiveGaugeVar}) as in 
(\ref{fix-u-}).
If we also represent the Lagrangian in the form 
\be
L^{++}(z,u^+)={\rm i} \,u^{+\1}u^{+\2}L(z,\z)
= {\rm i} (u^{+\1})^2 \,\z\,L(z,\z)~,
\label{L++-->L}
\ee
the action reduces to 
\bea
S= \frac{1}{ 16}\oint  \frac{\rd\z }{ 2\pi\ri}
\int\rd^4 x\,\z\,
({D}^{\1})^2({\bar D}_{\2})^2L(z,\z)\Big| \! \Big|~.
\label{ac2}
\eea
${}$Finally, making use of the analyticity of $L$ gives 
\bea
S=
 \frac{1 }{ 2\pi\ri }
 \oint 
 \frac{\rd\z }{  \z}
\int\rd^4 x\,{\rm d}^4\q \,
L(z,\z)\Big|~,
\eea
where the integration is carried out over the $\cN=1$ superspace.

The Lagrangian $L(z,\z)$ introduced in (\ref{L++-->L}) is characterized by
 the following superconformal transformation:
\bea
- \z\,\d L = 
\pa_a  \Big( \x^a\, \z \,L\Big)
+ D^-_\a  \Big( \x^{+\a}\, \z \,L\Big)  + {\bar D}^-_{\dt \a}\Big( {\bar \x}^{+\dt \a} \,\z\,L \Big)
+\pa_\z \Big(\L^{++}(\z) \z\,  L \Big) ~. 
\label{volume3}
\eea
It makes obvious the superconformal invariance of (\ref{ac2}).
Eq. (\ref{volume3})
can be compared with the five-dimensional transformation in \cite{K}.

 \sect{4D \mbox{$\bm{ \cN=2}$} superconformal theories}
 
 \subsection{Superconformal tensor and  \mbox{$\bm{O(2n)}$} multiplets}
 Superconformal self-couplings of tensor multiplets are well-known
 \cite{KLR,deWRV}.
 ${}$For  a set of tensor multiplets $H^{++I}  $,  with $I=1,\dots, n$, 
 superconformal dynamics is generated by 
 a Lagrangian $L^{++}  = L \big(H^{++I} \big)$
that is a real homogeneous function 
of first degree in the variables $H^{++}$, 
\bea
H^{++I} \frac{\pa}{\pa H^{++I} } \,L \big(H^{++} \big) 
= L \big(H^{++}\big)~.
\eea
Generalizations for $O(2n)$ multilets are obvious. 

To describe the improved $\cN=2$ tensor multiplet
\cite{improved} in projective superspace, 
some special considerations are required. 
But since such a formulation is well-known  \cite{KLR,HitKLR,deWRV}, 
we will not discuss it here.

We should point out that some examples
of superconformal self-couplings for tensor and $O(4)$ 
multiplets in harmonic superspace were given in \cite{GIO} and
\cite{Ketov} respectively.

 \subsection{Superconformal polar multiplets}
 
We consider a system of interacting arctic weight-one multiplets 
$\U^{+ } (z,u^+) $ and their smile-conjugates
$ \widetilde{\U}^{+}$ described by the Lagrangian \cite{K,KT-M}
\bea
L^{++} = {\rm i} \, K(\U^+, \widetilde{\U}^+)~,
\label{conformal-sm}
\eea
with $K(\F^I, {\bar \F}^{\bar J}) $ a real analytic function
of $n$ complex variables $\F^I$, where $I=1,\dots, n$.
Since $L^{++} =L^{++} (z,u^+) $ is required to  be a weight-two projective 
superfield, the potential  $K$ has to respect the following homogeneity condition
\be
\Big( \F^I \frac{\pa}{\pa \F^I} +  {\bar \F}^{\bar I} \frac{\pa}{\pa {\bar \F}^{\bar I}} \Big)
K(\F, \bar \F) = 2\, K( \F,   \bar \F) ~.
\label{Kkahler}
\ee
${}$For $L^{++}$ to be real, we require a stronger condition
\bea
 \F^I \frac{\pa}{\pa \F^I} K(\F, \bar \F) =  K( \F,   \bar \F)~.
 \label{Kkahler2}
 \eea
Then, representing $\U^+(z,u^+) =u^{+\1} \,\U(z,\z) $
and $\widetilde{\U}^+(z,u^+) =u^{+\2} \,\widetilde{\U}(z,\z) $, we 
can rewrite the Lagrangian in the form
\bea
L^{++} (z,u^+)= {\rm i} \, u^{+\1} u^{+\2} \, L(z,\z) ~, \qquad 
L =K(\U, \widetilde{\U})~.
\eea
The action takes the form
\bea
S=
 \frac{1 }{ 2\pi\ri }
 \oint 
\frac{\rd\z }{  \z}
\int \rd^4 x\,{\rm d}^4\q \,
K(\U^I, \widetilde{\U}^{\bar J})~,
\label{conformal-sm2}
\eea
with the integration contour around the origin in $\mathbb C$.
We should emphasise that action (\ref{conformal-sm2}) 
is formulated in terms of $\cN=1$ superfields,
but it is invariant under linearly realized $\cN=2$
superconformal transformations.

There  is a simple algebraic construction 
to generate superconformal actions of the form (\ref{conformal-sm2}). 
Let $\cP_n(w^a) = \cP_n(w^1, \dots, w^q)$ be a homogeneous  polynomial 
of order $n$ in 
$q$ complex variables $w^a$, $\cP_n(c \, w^a) = c^n\, \cP_n(w^a)$. 
Given a constant Hermitian matrix $\eta_{ {\bar a} b}$, we consider the action 
\bea
S=
 \frac{1 }{ 2\pi\ri }
 \oint 
\frac{\rd\z }{  \z}
\int \rd^4 x\,{\rm d}^4\q \,
\widetilde{\U}^{\bar a}  \,  \eta_{{\bar a} b} \,  \U^b~,
\label{conformal-sm3}
\eea
with the weight-one arctic multiplets $\U^a (z,\z)$ obeying the constraint 
\be
\cP_n(\U^a) =0~.
\ee

Suppose that the dynamical variables $\U^I (z, \z)$ in (\ref{conformal-sm2})
include a compensator $\U(z,\z)$, that is an arctic multiplet such that its 
lowest-order ($\z$-independent) 
component $\U_0$ is everywhere non-vanishing. 
Then, we can introduce new dynamical variables comprising the unique 
weight-one multiplet $\U(z,\z)$ and some set of weight-zero arctic multiplets
$\u^i(z,\z)$. The action (\ref{conformal-sm2}) will then turn into 
\bea
S=
 \frac{1 }{ 2\pi\ri }
 \oint \frac{\rd\z }{  \z}
\int \rd^4 x\,{\rm d}^4\q \,
\widetilde{\U} \U \,{\rm e}^{\cK(\u, \widetilde{\u} )}~, 
\label{conformal-sm4}
\eea
with $\cK(\u, \widetilde{\u} )$ a K\"ahler potential. 
This action is invariant under K\"ahler tansformations 
\bea
\U ~\longrightarrow ~{\rm e}^{-\L(\u)} \, \U~, \qquad 
\cK(\u, \widetilde{\u} ) ~\to ~ \cK(\u, \widetilde{\u} )+ \L(\u) + {\bar \L}( \widetilde{\u})~,
\eea
with $\L $ a holomorphic function.
Action (\ref{conformal-sm4}) is reminiscent of that describing  a general chiral 
sigma-model in 4D $\cN=1$ old minimal supergravity provided  one switches off
the gravitational superfield $H^m$ and keeps only the chiral compensator $\vf$ alive
(see, e. g., \cite{BK} for a review), 
with the latter being replaced with  $\U$ in the $\cN=2$ case.

 \sect{Reduction to  \mbox{$\bm{ \cN=1}$} superfields}
 
 The important powerful feature of the projective supermultiplets 
is that  they admit a simple decomposition in terms of standard $\cN=1$ 
superfields. In the superconformal case, it is therefore useful 
to reduce the $\cN=2$ superconformal transformation laws 
of the projective supermultiplets to $\cN=1$ superfields. 
This is explicitly carried out in the present section. 
   
 \subsection{\mbox{$\bm{ \cN=1}$}  decomposition of  \mbox{$\bm{ \cN=2}$} 
 superconformal Killings}

It turns out that the $\cN=2$ superconformal Killing vector $\x$ 
generates three types of transformations
at the level of  $\cN=1$ superfields.  In terms of the $\cN=1$ projection
\bea
\x \big| &:= &\x^A\big| D_A~, \non \\
\big[ \x \big|\;,\; D^{\1}_\a \big] &=& { \o}_\a{}^\b \big| D^{\1}_\b 
- \Big(\bar{ {\s}}\big|  +\L_{\1}{}^{\1} \big|\Big)  D^{\1}_\a
- {\L}_{\2}{}^{\1} \big|\, D^{\2}_\a ~,
\label{31}
\eea
they are as follows:\\
${}\quad$ {\bf 1.} An arbitrary $\cN=1$ superconformal transformation generated by 
\bea
{\bm \x} = {\overline {\bm \x}} = {\bm \x}^a (z) \,\pa_a + {\bm \x}^\a (z)\,D_\a
+ {\bar {\bm \x}}_{\dt \a} (z)\, {\bar D}^{\dt \a}
\label{n=1scf1}
\eea  
such that 
\be
[{\bm \x} \;,\; D^i_\a ] 
= \bm { \o}_\a{}^\b  D_\b +
\Big({\bm \s} - 2 \bar{ {\bm \s}}  \Big) D_\a~, 
\label{n=1scf2}
\ee
see the appendix.
The components of $\bm \x$ and their descendants  $ \bm { \o}_\a{}^\b $ and $\bm \s$
correspond  to the following choice of the parameters  in (\ref{31}):
\be
\x \big| ={\bm \x}~, \quad { \o}_\a{}^\b \big|  =\bm { \o}_\a{}^\b ~, 
\quad \s \big|= \bm \s ~,\quad
 {\L}_{\1}{}^{\1} \big|={\bar {\bm \s}}-{\bm \s}~, \quad
 {\L}_{\2}{}^{\1} \big|=0~.
\label{n=1scf3}
\ee
${}\quad$ {\bf 2.} An extended  superconformal transformation
generated by 
\bea
\x \big| &=& \r^\a D^{\2}_\a +{\bar \r}_{\dt \a} {\bar D}^{\dt \a}_{\2}~, 
\qquad \x^\a_{\2}  \big|=\r^\a~,
\non \\
{ \o}_\a{}^\b \big|  &=& \s \big|=
 {\L}_{\1}{}^{\1} \big|=0~, \qquad
 {\L}_{\2}{}^{\1} \big|=  {\L}^{\1 \1} \big|=
- \hf D^\a \r_\a~.
\label{esc}
\eea
${}\quad$ {\bf 3.} A shadow chiral rotation.
This  is a phase transformation
 of $\q^\a_{\2}$ only, with $\q^\a_{\1}$ kept unchanged, 
and it corresponds to the choice 
\bea
\x \big| =0~, \qquad { \o}_\a{}^\b \big|  =  {\L}_{\2}{}^{\1} \big|=0~, \qquad 
 \s \big| = {\L}_{\1}{}^{\1} \big| = - \bar \s \big | =-\frac{\rm i}{2}\,\a~.
\label{shadow1}
\eea

The spinor parameter $\r^\a$ in (\ref{esc}) can be shown to obey the equations
\bea
{\bar D}_{\dt \a} \r^\b =0~, \qquad D^{(\a}\r^{\b )}=0~,
\eea
and the latter imply 
\be
\pa^{{\dt \a} (\a} \r^{\b )} = D^2 \r^\b =0~.
\ee
There are several ordinary (component) transformations generated by 
the chiral spinor $\r^\a$ in (\ref{esc}): (i) second Q-supersymmetry 
transformation $(\e^\a$); (ii) off--diagonal SU(2)-transformation 
($  {\l} =  {\L}^{\1 \1}|_{\q=0} $); (iii) second S-supersymmetry transformation 
(${\bar \eta}_{\dt \a}$). They emerge as follows: $\r^\a(x_{(+)}, \q) = \e^\a
+ \l\, \q^\a - {\rm i} \,{\bar \eta}_{\dt \a}\, x^{{\dt \a}\a}_{(+)}\,$, with $x_{(+)}^a$ 
the chiral extension of $x^a$.

\subsection{\mbox{$\bm{ \cN=1}$}  superconformal transformations} 
Let us first consider how the $\cN=2$ superconformal multiplets 
vary  under  the $\cN=1$ superconformal transformations described by eqs.
(\ref{n=1scf1} -- \ref{n=1scf3}).
Here the superconfomal building blocks (\ref{L++Sigma}) take the form:
\bea
\L^{++}(\z) \big|=2\z(\bar{\bm \s} - {\bm \s})~,\qquad 
\S(\z) \big|= 2{\bm \s}~.
\label{n=1scf4} 
\eea

Consider the arctic multiplet of weight $n$, eq. (\ref{arctic1}).
Its $\cN=2$ superconformal transformation law (\ref{arctic2}) implies
\bea
\d \U_k = -{\bm \x} \U_k - 2k(\bar{\bm \s} - {\bm \s})\U_k -2n {\bm \s}\U_k~.
\label{arctic3} 
\eea
In particular, for the leading chiral $\F:= \U_0$ and complex linear 
$\S:=\U_1$ components we get
\bea
\d \F = -{\bm \x} \F  -2n{\bm \s} \F~, \qquad 
\d \S = -{\bm \x} \S - 2\bar{\bm \s} \S
-2(n-1){\bm \s} \S~.
\label{arctic4} 
\eea
These transformation laws can be seen to be  consistent with the 
off-shell constraints ${\bar D}_{\dt \a} \F =0$ and ${\bar D}^2 \S =0$. 

Consider the real $O(2n)$ multiplet (\ref{o2n2}).
Its $\cN=2$ superconformal transformation law 
(\ref{o2n2}) implies
\bea
\d H_k = -{\bm \x} H_k 
+2(k-n){\bm \s} H_k
- 2(k+n)\bar{\bm \s} H_k~.
\label{o2n3} 
\eea
In particular, for the leading chiral $\F:= H_{-n}$ and complex linear 
$\S:=H_{-n+1}$ components we get
\bea
\d \F = -{\bm \x} \F  -4n{\bm \s} \F~, \qquad 
\d \S = -{\bm \x} \S - 2\bar{\bm \s} \S
-2(2n-1){\bm \s} \S~.
\label{o2n4} 
\eea
These transformation laws  are consistent with the 
off-shell constraints ${\bar D}_{\dt \a} \F =0$ and ${\bar D}^2 \S =0$. 
As is seen from (\ref{o2n3}), the variation of the real superfield $H_0$ is real.

${}$For $n>1$, the real $O(2n)$ multiplet describes an off-shell hypermultiplet.
The special case $n=1$ corresponds to an off-shell tensor multiplet.
In accordance with (\ref{o2n4}), the real linear superfield $G:=H_0=\bar G$
transforms as
\bea
\d G = -{\bm \x} G - 2(\bar{\bm \s} +{\bm \s}) G~.
\label{o2n5} 
\eea
This transformation law is uniquely fixed by the off-shell constraints
${\bar D}^2 G =D^2G=0$. 

\subsection{Extended  superconformal transformations} 
We now turn to the extended superconformal transformations
(\ref{esc}). In this case, the superconformal building blocks are
\bea
\L^{++}(\z) \big|=-\hf \big(\z^2 D^\a\r_\a +{\bar D}_{\dt \a} {\bar \r}^{\dt \a} \big)
~,\qquad 
\S(\z) \big|= \hf \z D^\a\r_\a~.
\label{esc2} 
\eea
To read off the corresponding transformations of the component 
$\cN=1$ superfields of $\cN=2$ multiplets, it remains to use the identity
\bea
\big( \r^\a D^{\2}_\a +{\bar \r}_{\dt \a} {\bar D}^{\dt \a}_{\2} \big) Q^{[n]}(z,\z)
= \big( \z \r^\a D_\a - \frac{1}{\z}{\bar \r}_{\dt \a} {\bar D}^{\dt \a} \big) Q^{[n]}(z,\z)
\eea
that follows form the analyticity constraint.

${}$For  the arctic multiplet of weight $n$, eq. (\ref{arctic1}), we obtain
\begin{subequations}
\bea
\d \U_0 &=& {\bar \r}_{\dt \a} {\bar D}^{\dt \a} \U_1 
+\hf \big( {\bar D}_{\dt \a} {\bar \r}^{\dt \a} \big) \U_1 ~,
\non \\
\d \U_1 &=&-\r^\a D_\a \U_0 
+  {\bar D}_{\dt \a}\big( {\bar \r}^{\dt \a}  \U_2 \big)
-\frac{n}{2}\big(D^\a\r_\a\big) \U_0 ~, 
\label{arctic5} \\
\d \U_k &=&-\r^\a D_\a \U_{k-1} 
+{\bar \r}_{\dt \a} {\bar D}^{\dt \a} \U_{k+1} \non \\
&&
+\hf (k-n-1)\big(D^\a\r_\a\big) \U_{k-1} 
+\hf (k+1) \big( {\bar D}_{\dt \a} {\bar \r}^{\dt \a} \big) \U_{k+1}
~, \qquad k>1~. 
\label{arctic6} 
\eea
\end{subequations}
One can see that the transformation laws in (\ref{arctic5}) 
are consistent  with the 
off-shell constraints ${\bar D}_{\dt \a} \U_0 =0$ and ${\bar D}^2 \U_1 =0$. 

${}$For the real $O(2n)$ multiplet (\ref{o2n2}), we obtain
\begin{subequations}
\bea
\d H_{-n} &=& {\bar \r}_{\dt \a} {\bar D}^{\dt \a} H_{-n+1}
+\hf \big( {\bar D}_{\dt \a} {\bar \r}^{\dt \a} \big) H_{-n+1} ~,
\non \\
\d H_{-n+1} &=&-\r^\a D_\a H_{-n}
+  {\bar D}_{\dt \a} \big({\bar \r}^{\dt \a}  H_{-n+2} \big)
-{n}\big(D^\a\r_\a\big) H_{-n} ~,
\label{o2n6}  \\
\d H_k &=&-\r^\a D_\a H_{k-1} 
+{\bar \r}_{\dt \a} {\bar D}^{\dt \a} H_{k+1} 
-\hf (n+1-k)(D^\a\r_\a) H_{k-1} \non \\
&&
+\hf (n+1+k) \big( {\bar D}_{\dt \a} {\bar \r}^{\dt \a} \big) H_{k+1}~,
\qquad -n+1<k<0~, 
\label{o2n7} \\
\d H_0 &=&-\r^\a D_\a H_{-1} 
-{\bar \r}_{\dt \a} {\bar D}^{\dt \a} \bar{H}_{-1} 
-\hf (n+1)\Big((D^\a\r_\a) H_{-1} 
+\big( {\bar D}_{\dt \a} {\bar \r}^{\dt \a} \big) 
\bar{H}_{-1} \Big)~.~~~~~~
\label{o2n8} 
\eea
\end{subequations}

\subsection{Shadow chiral rotation}
${}$Finally, let us consider the shadow chiral rotation
(\ref{shadow1}).
In the case of the arctic multiplet of weight $n$, eq. (\ref{arctic1}),
it acts as follows:
\be
\d \U_k = {\rm i}\,\a (k-\frac{n}{2}) \U_k~.
\label{shadow 2}
\ee
${}$For the real $O(2n)$ multiplet (\ref{o2n2}), we obtain
\be
\d H_k = {\rm i}\,\a kH_k~.
\label{shadow3}
\ee
The component $H_0$ is real, and therefore it does not transform.
In a finite form, this transformation reads 
\bea
\U(z,\z) \quad &\longrightarrow & \quad \U'(z,\z)=
{\rm e}^{-{\rm i} (n/2) \a}\, \U(z,{\rm e}^{{\rm i} \a} \z)~, 
\label{shadow4}\\
H^{[2n]}(z,\z) \quad &\longrightarrow & \quad 
H^{[2n]}{}'(z,\z)=
 H^{[2n]}(z,{\rm e}^{{\rm i} \a} \z)~.
\label{shadow5}
\eea
 
\sect{Non-superconformal case: \mbox{$\bm{ \cN=2}$}  sigma-models 
on tangent bundles of K\"ahler manifolds}
Before turning to a  analysis of  the superconformal 
dynamical system  (\ref{conformal-sm2}),
it is instructive 
to consider a more general family of  4D $\cN=2$ off-shell 
supersymmetric nonlinear sigma-models that are described in 
ordinary $\cN=1$ 
superspace by the action\footnote{The study of such models 
was initiated in \cite{K98,GK1,GK2}, and important 
results have recently been obtained in \cite{AKL,AKL2}. 
They correspond to a subclass of the general hypermultiplet theories
in projective superspace \cite{LR1,LR2} 
obtained by replacing 
$K \big( \U , \widetilde{\U}   \big) 
\to K \big( \U , \widetilde{\U} , \z  \big) $
in (\ref{nact}). }
\bea
S[\U, \widetilde{\U}]  =  
\frac{1}{2\pi {\rm i}} \, \oint \frac{{\rm d}\z}{\z} \,  
 \int 
 \rd^4 x\,{\rm d}^4\q
 \, 
K \big( \U^I (\z), \widetilde{\U}^{\bar{J}} (\z)  \big) ~.
\label{nact} 
\eea
The arctic $\U (\z)$ and  antarctic $\widetilde{\U} (\z)$ dynamical variables  
are generated by an infinite set of ordinary superfields:
\be
 \U (\z) = \sum_{n=0}^{\infty}  \, \U_n \z^n = 
\F + \S \,\z+ O(\z^2) ~,\qquad
\widetilde{\U} (\z) = \sum_{n=0}^{\infty}  \, {\bar
\U}_n
 (-\z)^{-n}~.
\label{exp}
\ee
Here $\F$ is chiral, $\S$  complex linear, 
\be
{\bar D}_{\dt{\a}} \F =0~, \qquad \qquad {\bar D}^2 \S = 0 ~,
\label{chiral+linear}
\ee
and the remaining component superfields are unconstrained complex 
superfields.  
The above  theory
 occurs as a minimal $\cN=2$ extension of the
general four-dimensional $\cN=1$ supersymmetric 
nonlinear sigma-model \cite{Zumino}
\be
S[\F, \bar \F] =  \int 
\rd^4 x\,{\rm d}^4\q
\, K(\Phi^{I},
 {\bar \Phi}{}^{\bar{J}})  ~,
\label{nact4}
\ee
with $K$ the  K\"ahler potential of a K\"ahler manifold $\cM$.

The reason we are interested here in the  $\cN=2$ supersymmetric theory 
(\ref{nact}) is that its action  becomes superconformal upon imposing 
the homogeneity condition  (\ref{Kkahler2}).

\subsection{Background material on  \mbox{$\bm{ \cN=2}$}  sigma-models} 
The extended supersymmetric  sigma-model  (\ref{nact}) 
inherits  all the geometric features of
its $\cN=1$ predecessor (\ref{nact4}). 
The K\"ahler invariance of the latter,
$K(\F, \bar \F) \to K(\F, \bar \F) +
\L(\F)+  {\bar \L} (\bar \F) $,
turns into 
\be
K(\U, \widetilde{\U})  \quad \longrightarrow \quad K(\U, \widetilde{\U}) ~+~
\L(\U) \,+\, {\bar \L} (\widetilde{\U} ) 
\label{kahl2}
\ee
for the model (\ref{nact}).\footnote{In the superconfomal case,
the Lagrangian obeys the homogeneity condition  (\ref{Kkahler2}),
and no K\"ahler invariance survives.}
A holomorphic reparametrization of the K\"ahler manifold,
$ \F^I  \to \F'{}^I=f^I \big( \F \big) $,
has the following
counterpart
\be
\U^I (\z) \quad  \longrightarrow  \quad \U'{}^I(\z)=f^I \big (\U(\z) \big)
\label{kahl3}
\ee
in the $\cN=2$ case. Therefore, the physical
superfields of the 
${\cal N}=2$ theory
\be
 \U^I (\z)\Big|_{\z=0} ~=~ \F^I ~,\qquad  \quad \frac{ {\rm d} \U^I (\z) 
}{ {\rm d} \z} \Big|_{\z=0} ~=~ \S^I ~,
\label{kahl4} 
\ee
should be regarded, respectively, as  coordinates of a point in the K\" ahler
manifold and a tangent vector at  the same point. 
Thus the variables $(\F^I, \S^J)$ parametrize the tangent 
bundle $T\cM$ of the K\"ahler manifold $\cM$ \cite{K}. 

To describe the theory in terms of 
the physical superfields $\F$ and $\S$ only, 
all the auxiliary 
superfields have to be eliminated  with the aid of the 
corresponding algebraic equations of motion
\bea
\oint \frac{{\rm d} \z}{\z} \,\z^n \, \frac{\pa K(\U, \widetilde{\U} ) }{\pa \U^I} 
~ = ~ \oint \frac{{\rm d} \z}{\z} \,\z^{-n} \, \frac{\pa 
K(\U, \widetilde{\U} ) } {\pa \widetilde{\U}^{\bar J} } 
~ = ~ 0 ~, \qquad n \geq 2 ~ .               
\label{asfem}
\eea
Let $\U_*(\z) \equiv \U_*( \z; \F, {\bar \F}, \S, \bar \S )$ 
denote a unique solution subject to the initial conditions
\bea
\U_* (0)  = \F ~,\qquad  \quad \dt{\U}_* (0) 
 = \S ~.
\label{geo3} 
\eea

For a general K\"ahler manifold $\cM$, 
the auxiliary superfields $\U_2, \U_3, \dots$, and their
conjugates,  can be eliminated  only perturbatively. 
Their elimination  can be carried out
using the ansatz \cite{KL}
\bea
\U^I_n = 
\sum_{p=0}^{\infty} 
G^I{}_{J_1 \dots J_{n+p} \, \bar{L}_1 \dots  \bar{L}_p} (\F, {\bar \F})\,
\S^{J_1} \dots \S^{J_{n+p}} \,
{\bar \S}^{ {\bar L}_1 } \dots {\bar \S}^{ {\bar L}_p }~, 
\qquad n\geq 2~.
\label{ansatz}
\eea
Assuming that 
the auxiliary superfields 
have been eliminated, 
the action (\ref{nact}) should take the form\footnote{As compared with
the expressions in  \cite{GK1,GK2}, the series for $\cL$ contains 
an extra factor of $(-1)^n$. The reason for its insertion will become clear
in next subsection.} 
 \cite{GK1,GK2}:
\bea
S_{{\rm tb}}[\F,  \S]  &=&  
\frac{1}{2\pi {\rm i}} \, \oint \frac{{\rm d}\z}{\z} \,  
 \int \rd^4 x\,{\rm d}^4\q \, 
K \big( \U_* (\z), \breve{\U}_* (\z)  \big) \non \\
&=& \int 
\rd^4 x\,{\rm d}^4\q
\, \Big\{\,
K \big( \F, \bar{\F} \big)+  
\cL \big(\F, \bar \F, \S , \bar \S \big)\Big\}~,\non \\
\cL 
&=&
\sum_{n=1}^{\infty} (-1)^n \cL_{I_1 \cdots I_n {\bar J}_1 \cdots {\bar 
J}_n }  \big( \F, \bar{\F} \big) \S^{I_1} \dots \S^{I_n} 
{\bar \S}^{ {\bar J}_1 } \dots {\bar \S}^{ {\bar J}_n }
:= \sum_{n=1}^{\infty} (-1)^n \cL^{(n)} ~,~~~~~~~~
\label{act-tab}
\eea
where $\cL_{I {\bar J} }=   g_{I \bar{J}} \big( \F, \bar{\F}  \big) $ 
and the series coefficients $\cL_{I_1 \cdots I_n {\bar J}_1 \cdots {\bar 
J}_n }$, for  $n>1$, 
are tensor functions of the K\"ahler metric
$g_{I \bar{J}} \big( \F, \bar{\F}  \big) 
= \pa_I 
\pa_ {\bar J}K ( \F , \bar{\F} )$,  the Riemann curvature $R_{I {\bar 
J} K {\bar L}} \big( \F, \bar{\F} \big) $ and its covariant 
derivatives.  Each term in the action contains equal powers
of $\S$ and $\bar \S$, since the original model (\ref{nact}) 
is invariant under rigid U(1)  transformations\footnote{Transformation
(\ref{rfiber}) coincides with the shadow chiral rotation 
(\ref{shadow4}) for $n=0$.}
 \cite{GK1}
\be
\U(\zeta) ~~ \mapsto ~~ \U({\rm e}^{{\rm i} \a} \zeta) 
\quad \Longleftrightarrow \quad 
\U_n(z) ~~ \mapsto ~~ {\rm e}^{{\rm i} n \a} \U_n(z) ~.
\label{rfiber}
\ee

\subsection{Putting the extended supersymmetry to work}

In the  recent work \cite{AKL2}, it was demonstrated that 
supersymmetry considerations allow one to avoid 
the problem of solving the auxiliary field equations (\ref{asfem}) 
in the case of Hermitian symmetric spaces which possess
a covariantly constant curvature tensor. 
\be
\nabla_L  R_{I_1 {\bar  J}_1 I_2 {\bar J}_2}
= {\bar \nabla}_{\bar L} R_{I_1 {\bar  J}_1 I_2 {\bar J}_2} =0~.
\label{cc}
\ee
Here we address
the general case of an arbitrary K\"ahler manifold, with no pretense 
of completeness. 

The  theory under consideration,  eq. (\ref{nact}),  is $\cN=2$ super-Poincar\'e invariant. 
In terms of the superconformal formalism presented in section 2,
its symmetry structure is described by those transformations which 
are characterised by 
\bea
\L^{ij} = \s=0~.
\eea
These conditions correspond to the $\cN=2$ Killing supervectors.
In particular, the parameter $\r^\a$ in (\ref{esc}) should be restricted to be  a constant spinor,
$\r^\a =\ve^\a={\rm const}$. Then, the arctic multiplet transformation laws (\ref{arctic5}) and 
 (\ref{arctic6}) become 
 \begin{subequations}
\bea
\d \U_0 &=& {\bar \ve}_{\dt \a} {\bar D}^{\dt \a} \U_1 ~,
\qquad 
\d \U_1 =-\ve^\a D_\a \U_0 
+   {\bar \ve}_{\dt \a}{\bar D}^{\dt \a} \U_2 
~,  \label{arctic7} \\
\d \U_k &=&-\ve^\a D_\a \U_{k-1} 
+{\bar \ve}_{\dt \a} {\bar D}^{\dt \a} \U_{k+1} 
~, \qquad k>1~. 
\label{arctic8} 
\eea
\end{subequations}
Upon elimination of the auxiliary superfields, 
the action (\ref{act-tab}) should be invariant under
the supersymmetry transformations
\bea
\d \F &=& {\bar \ve}_{\dt \a} {\bar D}^{\dt \a} \S ~,
\qquad 
\d \S =-\ve^\a D_\a \F 
+   {\bar \ve}_{\dt \a}{\bar D}^{\dt \a} \U_2 \big(\F, \bar \F, \S , \bar \S \big)
~,  
\label{arctic9} 
\eea
where $\U_2$ now a composite field of the general form given in (\ref{ansatz}).
Since $\U_2$ transforms as a connection under 
the holomorphic reparametrizations (\ref{kahl3})
\bea
\U_2^I  \quad  \longrightarrow  \quad 
\U'{}^I_2 
= \hf  \frac{ \pa^2 f^I  \big( \F \big) }{\pa \F^J \pa \F^K}\, \S^J \S^K
+\frac{ \pa f^I  \big( \F \big) }{\pa \F^J }\, \S^J ~,
\eea
we can rewrite $\U_2$ in a slightly more specific form:
\bea
\U^I_2 &=& -\hf \G^I_{JK} \big( \F, \bar{\F} \big) \, \S^J\S^K+
\sum_{p=1}^{\infty} 
G^I{}_{J_1 \dots J_{p+2} \, \bar{L}_1 \dots  \bar{L}_p} (\F, {\bar \F})\,
\S^{J_1} \dots \S^{J_{p+2}} \,
{\bar \S}^{ {\bar L}_1 } \dots {\bar \S}^{ {\bar L}_p }~,~~~~~ \non \\
&:=& -\hf \G^I_{JK} \big( \F, \bar{\F} \big) \, \S^J\S^K+
\sum_{p=1}^{\infty} G^I_{(p)}~,
\label{ansatz2}
\eea
with $\G^I_{JK} 
( \F , \bar{\F} )$  the Christoffel symbols for the  
K\"ahler metric $g_{I \bar J} ( \F , \bar{\F} )$. 
Here the coefficients $G^I{}_{J_1 \dots J_{p+2} \, \bar{L}_1 \dots  \bar{L}_p} (\F, {\bar \F})$
are tensor functions of the K\"ahler metric,
the Riemann curvature $R_{I {\bar 
J} K {\bar L}} \big( \F, \bar{\F} \big) $ and its covariant 
derivatives.

Of course, the tensor fields 
$\cL_{I_1 \cdots I_n {\bar J}_1 \cdots {\bar J}_n }$ in (\ref{act-tab}) and 
$G^I{}_{J_1 \dots J_{p+2} \, \bar{L}_1 \dots  \bar{L}_p} $ in 
(\ref{ansatz2}) are uniquely determined, 
in the theory with action  (\ref{nact}), 
once (i) we have solved the auxiliary field equations (\ref{asfem}); 
and (ii) have done the contour integral in the first line of (\ref{act-tab}).
However, these two problems are tremendous in general.
There is an alternative approach.
We can look for a $\cN=1$ supersymmetric theory of the form  (\ref{act-tab}),
which is required to be invariant under extended supersymmetric transformations  (\ref{arctic9})
such that $\U^I_2 $ is of the general form (\ref{ansatz2}).
It is clear, from the previous considerations,  
that the requirement of extended supersymmetry 
should uniquely determine both sets of the coefficient functions 
$\cL_{I_1 \cdots I_n {\bar J}_1 \cdots {\bar J}_n }$  and 
$G^I{}_{J_1 \dots J_{p+2} \, \bar{L}_1 \dots  \bar{L}_p} $.
And it does indeed, as can be explicitly checked in  leading orders 
of perturbation theory.
Here are some low-order results:
\begin{subequations}
\bea
\cL^{(1)} &=&   g_{I \bar{J}} \S^I {\bar \S}^{\bar J}~, 
\label{L1} \\
\cL^{(2)} &=& \frac{1}{4} R_{I_1 {\bar J}_1 I_2 {\bar J}_2} \S^{I_1}\S^{I_2}
{\bar \S}^{{\bar J}_1}{\bar \S}^{{\bar J}_2}~,
\label{L2}  \\
\cL^{(3)} &=& \frac{1}{12} \Big\{ \frac{1}{6} 
\{ \nabla_{I_3}, {\bar \nabla}_{{\bar J}_3} \}
R_{I_1 {\bar J}_1 I_2 {\bar J}_2} 
+R_{I_1 {\bar J}_1 I_2 }{}^LR_{L {\bar J}_2 I_3 {\bar J}_3}\Big\}  
\S^{I_1}\dots \S^{I_3}
{\bar \S}^{{\bar J}_1}\dots{\bar \S}^{{\bar J}_3}~,~~~~~~
\label{L3}
\eea
\end{subequations}
and 
\bea
G^L_{(1)} = \frac{1}{6} \nabla_{I_3}R_{I_1 {\bar J} I_2 }{}^L \,
 \S^{I_1}\dots \S^{I_3}{\bar \S}^{{\bar J}}~,
\eea
The expressions for $\cL^{(1)}$ and $\cL^{(2)}$ first appeared in \cite{K} and 
\cite{GK1} respectively.

Before continuing on, we should recall the important notion 
of {\it canonical} coordinate system for K\"ahler manifolds
that was introduced by Bochner in 1947 \cite{Bochner} 
and later used by Calabi in the 1950s \cite{Calabi}.\footnote{This
coordinate system was  re-discovered by supersymmetry  practitioners
in the 1980s under the name {\it normal gauge} \cite{GGRS,A-GG,HKLR}.}
In a neighborhood of
 any point $p$ of the  K\"ahler manifold $\cM$,  
holomorphic reparametrizations  and K\"ahler transformations
can be used to choose a coordinate system, with origin at $p$,
in which the K\"ahler potential takes the form:  
\bea
\bm{K} (\f, \bar \f ) &=&{\bm g}_{I \bar{J}} \,\f^I {\bar \f}^{\bar J}
+ \sum_{m,n \geq 2}^{\infty}  {\bm K}^{(m,n)} (\f, \bar \f)~,
\non \\
{\bm K}^{(m,n)} (\f, \bar \f) &:=&
\frac{1}{m! n!}\,
{\bm K}_{I_1 \cdots I_m {\bar J}_1 \cdots {\bar 
J}_n }  \, \f^{I_1} \dots \f^{I_m} 
{\bar \f}^{ {\bar J}_1 } \dots {\bar \f}^{ {\bar J}_n }~.
\label{normal-gauge} 
\eea
In such a coordinate system, there still remains the freedom to perform 
linear holomorphic reparametrizations which can be used 
to set the metric at the origin, $p \in \cM$, to  be ${\bm g}_{I \bar{J}}= \d_{I \bar{J}}$.
The Taylor coefficients, ${\bm K}_{I_1 \cdots I_m {\bar J}_1 \cdots {\bar J}_n } $,
 in (\ref{normal-gauge}) turn out to be tensor functions of the K\"ahler metric,
the Riemann curvature $R_{I {\bar J} K {\bar L}}  $ and its covariant 
derivatives,  all of them  evaluated at the origin.
In particular, one finds\footnote{These results are easily derived by applying  the
relation $  K_{I_1 I_2{\bar J}_1 {\bar J}_2}= R_{I_1 {\bar J}_1 I_2 {\bar J}_2}
+g_{M \bar N} \G^M_{I_1I_2} {\bar \G}^{\bar N}_{{\bar J}_1{\bar J}_2} $.}
\begin{subequations}
\bea
{\bm K}^{(2,2)} &=& \frac{1}{4} R_{I_1 {\bar J}_1 I_2 {\bar J}_2}\, \f^{I_1}\f^{I_2}
{\bar \f}^{{\bar J}_1}{\bar \f}^{{\bar J}_2}~,
\label{(2,2)} \\
{\bm K}^{(3,2)} &=& \frac{1}{12} 
\nabla_{I_3} R_{I_1 {\bar J}_1 I_2 {\bar J}_2} \,
\f^{I_1} \f^{I_2} \f^{I_3}
{\bar \f}^{{\bar J}_1}{\bar \f}^{{\bar J}_2}~, 
\label{(3,2)}\\
{\bm K}^{(4,2)} &=& \frac{1}{48} 
\nabla_{I_3} \nabla_{I_4} R_{I_1 {\bar J}_1 I_2 {\bar J}_2} \,
\f^{I_1} \dots \f^{I_4}
{\bar \f}^{{\bar J}_1}{\bar \f}^{{\bar J}_2}~, 
\label{(4,2) } \\
{\bm K}^{(3,3)} &=& \frac{1}{12} \Big\{ \frac{1}{6} 
\{ \nabla_{I_3}, {\bar \nabla}_{{\bar J}_3} \}
R_{I_1 {\bar J}_1 I_2 {\bar J}_2} 
+R_{I_1 {\bar J}_1 I_2 }{}^LR_{L {\bar J}_2 I_3 {\bar J}_3}\Big\}  
\f^{I_1} \dots \f^{I_3}
{\bar \f}^{{\bar J}_1}\dots{\bar \f}^{{\bar J}_3}~,~~~~~~
\label{(3,3)}  \\
{\bm K}^{(4,3)} &=& \frac{1}{144} \Big\{  
{\bar \nabla}_{{\bar J}_3} \nabla_{I_3} \nabla_{I_4} 
R_{I_1 {\bar J}_1 I_2 {\bar J}_2} 
+6 R_{I_3 {\bar J}_3 I_4 }{}^L
\nabla_{L} R_{I_1 {\bar J}_1 I_2 {\bar J}_2} \non \\  
&& \qquad \qquad \qquad + 4 R_{I_1 {\bar J}_1 L {\bar J}_2}
\nabla_{I_2} R_{I_3 {\bar J}_3 I_4}{}^L \Big\}
\f^{I_1} \dots \f^{I_4}
{\bar \f}^{{\bar J}_1}\dots{\bar \f}^{{\bar J}_3} \non \\
&= &  \frac{1}{144} 
{\bm K}_{I_1 \cdots I_4 {\bar J}_1 \cdots {\bar J}_3 }  \,
\f^{I_1} \dots \f^{I_4}
{\bar \f}^{{\bar J}_1}\dots{\bar \f}^{{\bar J}_3}~,
\label{(4,3)} 
\eea
\end{subequations}
\addtocounter{equation}{-1}
\begin{subequations}
\setcounter{equation}{5}
\bea
{\bm K}^{(4,4)} &=& \frac{1}{576} \Big\{  
{\bar \nabla}_{{\bar J}_4}  {\bm K}_{I_1 \cdots I_4 {\bar J}_1 \cdots {\bar J}_3 } 
+6 R_{I_3 {\bar J}_3 I_4 }{}^L {\bar \nabla}_{{\bar J}_4} 
\nabla_{L} R_{I_1 {\bar J}_1 I_2 {\bar J}_2} \non \\
&&+ 4 \big( {\bar \nabla}_{{\bar J}_4}  R_{I_1 {\bar J}_1 L {\bar J}_2} \big)
\nabla_{I_2} R_{I_3 {\bar J}_3 I_4}{}^L \non \\
&&+6 R_{I_1 {\bar J}_1 I_2}{}^K\Big(R_{I_3 {\bar J}_2 I_4}{}^L
 R_{K {\bar J}_3 L {\bar J}_4} 
+2R_{I_3 {\bar J}_2 K}{}^L R_{I_4 {\bar J}_3 L {\bar J}_4} 
\Big) \Big\} \f^{I_1} \dots \f^{I_4}
{\bar \f}^{{\bar J}_1}\dots{\bar \f}^{{\bar J}_4} 
\non \\
&= &  \frac{1}{576} 
{\bm K}_{I_1 \cdots I_4 {\bar J}_1 \cdots {\bar J}_4 }  \,
\f^{I_1} \dots \f^{I_4}
{\bar \f}^{{\bar J}_1}\dots{\bar \f}^{{\bar J}_4}~.
\label{(4,4)}
\eea
\end{subequations}
It is possible to rewrite ${\bm K}^{(4,4)} $ in a manifestly real form, but such an expression 
appears to be   much longer than (\ref{(4,4)}).
The relations (\ref{(2,2)}--\ref{(3,3)}) appeared earlier in \cite{HIN}.

We should point out that in the literature, 
there exist closed-form   expressions \cite{normal}
for the Riemann normal coordinate expansion.
It would be very interesting to obtain a similar expression for the canonical 
coordinate system.

The above relations hint at the fact that, for $m\neq n$,  
the tensor ${\bm K}_{I_1 \cdots I_m {\bar J}_1 \cdots {\bar J}_n }$ 
should be a sum of terms each of which is
proportional to  a (multiple) covariant derivative of the Riemann tensor. 
In other words, 
\bea
\nabla_L  R_{I_1 {\bar  J}_1 I_2 {\bar J}_2}
= {\bar \nabla}_{\bar L} R_{I_1 {\bar  J}_1 I_2 {\bar J}_2} =0
\qquad \Longrightarrow \qquad 
{\bm K}^{(m,n)}=0~, \quad m\neq n~.
\label{covar-const}
\eea
Indeed, this holds in general.

If one compares the expressions for $\cL^{(2)}$ and  $\cL^{(3)}$, 
eqs. (\ref{L2}) and    (\ref{L3}),  with those for 
${\bm K}^{(2,2)}$ and ${\bm K}^{(3,3)}$ above, it is tempting to conclude that 
$$
\cL^{(n)} ~= ~{\bm K}^{(n,n)}(\f \to \S , \bar \f  \to \bar \S)~.
$$
Unfortunately, this does not hold in general, 
since for $n=4$ one finds 
\bea
\cL^{(4)} &=& \Big\{  \frac{1}{576} 
{\bm K}_{I_1 \cdots I_4 {\bar J}_1 \cdots {\bar J}_4 }  
-\frac{1}{36} \big( {\bar \nabla}_{{\bar J}_4}  R_{I_1 {\bar J}_1 L {\bar J}_2} \big)
\nabla_{I_2} R_{I_3 {\bar J}_3 I_4}{}^L \Big\}
\S^{I_1} \dots \S^{I_4}
{\bar \S}^{{\bar J}_1}\dots{\bar \S}^{{\bar J}_4}~,~~~~~~~~
\label{L4}
\eea
compare with (\ref{(4,4)}). However, the correct statement is the following:
\be
\cL^{(n)} ~= ~{\bm K}^{(n,n)}(\f \to \S , \bar \f  \to \bar \S)~+~
(\nabla R)\mbox{-terms}~.
\label{statement}
\ee
Here the second term on the right consists of those terms that vanish 
in the limit (\ref{cc}).
Eq. (\ref{statement}) is one of the main results of this work.

In deriving (\ref{L4}), one has to make use of the expression for 
$G^L_{(2)}$ that appears in (\ref{ansatz2}). 
It is 
\bea
G_{(2)\,{\bar J}_3} & \equiv  & g_{ {\bar J}_3 L} G^L_{(2)}
=G_{I_1 \dots I_{4} \, \bar{J}_1  \bar{J}_2; {\bar J}_3} \,
\S^{I_1} \dots \S^{I_{4}} \,
{\bar \S}^{ {\bar J}_1 }  {\bar \S}^{ {\bar J}_2 } \non \\
&=& \frac{1}{6} \Big\{ \big(\nabla_{I_4} R_{I_1 {\bar J}_1 I_2}{}^L\big)\, 
R_{L {\bar J}_2 I_3 {\bar J}_3} 
-\frac{1}{8} \bm{K}_{I_1 \dots I_{4} \, \bar{J}_1   \bar{J}_2 {\bar J}_3} \Big\} \,
\S^{I_1} \dots \S^{I_{4}} \,
{\bar \S}^{ {\bar J}_1 }  {\bar \S}^{ {\bar J}_2 } ~.~~~~
\eea
More generally, for any term in the series  in (\ref{ansatz2}),
it should hold
\bea
G_{(n)\,{\bar J}_{n+1}} ~ \equiv  ~ g_{ {\bar J}_{n+1} L} G^L_{(n)}
~\propto ~{\bm K}^{(n+2,n+1)}(\f \to \S , \bar \f  \to \bar \S)~+~
(\nabla R)\mbox{-terms}~.
\eea

Let us recall that $\U_*(\z) \equiv \U_*( \z; \F, {\bar \F}, \S, \bar \S )$ 
denotes the unique solution 
to the auxiliary field equations (\ref{asfem}) 
under the initial conditions (\ref{geo3}).
We conjecture that $\U_*(\z)$ 
obeys the following
generalised geodesic equation:
\bea
\frac{ {\rm d}^2 \U^I_* (\z) }{ {\rm d} \z^2 } &+& 
\G^I_{JK} \Big( \U_* (\z), \bar{\F} \Big)\,
\frac{ {\rm d} \U^J_* (\z) }{ {\rm d} \z } \,
\frac{ {\rm d} \U^K_* (\z) }{ {\rm d} \z }  \non \\
&=& 
2\sum_{p=1}^{\infty} 
G^I{}_{J_1 \dots J_{p+2} \, \bar{L}_1 \dots  \bar{L}_p} (\U_*, {\bar \F})\,
\frac{ {\rm d} \U^{J_1}_* (\z) }{ {\rm d} \z } \,
\dots 
\frac{ {\rm d} \U^{J_{p+2}}_* (\z) }{ {\rm d} \z } \,
{\bar \S}^{ {\bar L}_1 } \dots {\bar \S}^{ {\bar L}_p }~,
\label{gengeodesic}
\eea
and is its unique solution 
under the same initial conditions. 
This equation is covariant with respect to   holomorphic reparametrizations
of the K\"ahler manifold.
If the curvature tensor is covariantly constant, 
eq. (\ref{gengeodesic}) 
reduces to the geodesic equation given in \cite{GK1,GK2}.

At the moment, we do not know the explicit structure 
of the derivatives terms in (\ref{statement}). We believe that 
a more systematic analysis of the invariance under extended supersymmetry
transformations would allow one to determine these terms.

If the curvature tensor is covariantly constant, (\ref{cc}), 
there occur dramatic simplifications. In particular, here we obtain
\bea
\bm{K} (\f, \bar \f ) ={\bm g}_{I \bar{J}} \,\f^I {\bar \f}^{\bar J}
&+&
\sum_{n = 2}^{\infty}  
\frac{1}{(n!)^2}\,
{\bm K}_{I_1 \cdots I_n {\bar J}_1 \cdots {\bar 
J}_n }  \, \f^{I_1} \dots \f^{I_n} 
{\bar \f}^{ {\bar J}_1 } \dots {\bar \f}^{ {\bar J}_n }~, \non \\
\cL =
- g_{I \bar{J}} \S^I {\bar \S}^{\bar J}
&+&\sum_{n=2}^{\infty }
\frac{(-1)^n}{(n!)^2}\,
{\bm K}_{I_1 \cdots I_n {\bar J}_1 \cdots {\bar 
J}_n }  \, \S^{I_1} \dots \S^{I_n} 
{\bar \S}^{ {\bar J}_1 } \dots {\bar \S}^{ {\bar J}_n }~.
\label{sym}
\eea
In refs. \cite{AKL,AKL2}, the sigma-model (\ref{nact}) 
was explicitly `solved' for all Hermitian symmetric spaces except
${\rm E}_7/{\rm E}_6 \times {\rm U}(1)$. The above result 
allows one to address this case. 
Still, it would be very interesting to apply the scheme presented in 
\cite{AKL2} to the case of ${\rm E}_7/{\rm E}_6 \times {\rm U}(1)$. 

\sect{Back to the superconformal case}
${}$For the dynamical system (\ref{nact}), 
we have demonstrated that its description in terms of the physical 
superfields $(\F, \S)$, eq. (\ref{act-tab}), can be achieved 
by making use of the power of 
$\cN=2$ Poincar\'e supersymmetry, without the need to solve
the auxiliary field equations (\ref{asfem}).  
Now we are prepared to turn to the analysis 
of the general superconformal sigma-model (\ref{conformal-sm2}).

The action (\ref{nact})  becomes superconformal upon imposing 
the homogeneity condition  (\ref{Kkahler2}), and hence the symmetry 
group gets enhanced. In particular, the action (\ref{act-tab})
associated with (\ref{conformal-sm2}) should be invariant 
under $\cN=1$ superconformal transformations
\bea
\d \F = -{\bm \x} \F  -2{\bm \s} \F~, \qquad 
\d \S = -{\bm \x} \S - 2\bar{\bm \s} \S
\eea
and extended supeconformal transformations
\bea
\d \F &=& {\bar \r}_{\dt \a} {\bar D}^{\dt \a} \S 
+\hf \big( {\bar D}_{\dt \a} {\bar \r}^{\dt \a} \big) \S ~,
\non \\
\d \S &=&-\r^\a D_\a \F 
-\frac{n}{2}\big(D^\a\r_\a\big) \F
+  {\bar D}_{\dt \a}\Big\{ {\bar \r}^{\dt \a}  \U_2(\F, \bar \F, \S, \bar \S )  \Big\}
~,
\eea 
where $ \U_2(\F, \bar \F, \S, \bar \S ) $ is given by eq. (\ref{ansatz2}).
What are the implications of these additional symmetries?
Actually  it can be seen that no additional implications occur.
If the action (\ref{act-tab}) is $\cN=2$ supersymmetric, and the 
K\"ahler potential $K(\F ,\bar \F )$ obeys the homogeneity condition  (\ref{Kkahler2}), 
the theory is $\cN=2$ superconformal.
\\

While this paper was in the process of writing-up, 
there appeared a new work in the archive \cite{HS}, 
in which some superconformal aspects of 4D $\cN=2$ projective superspace 
were discussed, see also \cite{K}.
\\

\noindent
{\bf Acknowledgements:}\\
Discussions with Ulf Lindstr\"om, Martin Ro\v{c}ek 
and Gabriele Tartaglino-Mazzucchelli are 
gratefully acknowledged.
It is pleasure to acknowledge hospitality of the 
 5th Simons Workshop where this project was largely completed. 
This work is supported  in part by the Australian Research Council.

\appendix

\sect{\mbox{$\bm{ \cN}$}-extended superconformal Killing vectors}
\label{Killing}

In the main body of this paper, we have made extensive use of 
the $\cN=1$ and $\cN=2$ superconformal Killing vectors. 
Here we collect, following \cite{Park,KT},  
the essential information about the $\c$-extended superconformal 
Killing vectors, specifically for $\cN \leq 3$.

In 4D $\cN$-extended
superspace ${\mathbb R}^{4|4\cN}$ parametrized  
by  coordinates  $ z^A = (x^a,  \q^\a_i, {\bar \q}^i_\dt{\a} )$, 
with $i=1,\dots, \cN$,
an infinitesimal superconformal  transformation 
$ z^A \to    z^A  + \x \cdot z^A $ is generated by 
a superconformal  Killing vector
\be
\x = {\overline \x} = \x^a (z) \,\pa_a + \x^\a_i (z)\,D^i_\a
+ {\bar \x}_{\dt \a}^i (z)\, {\bar D}^{\dt \a}_i
\ee   
defined to satisfy 
\be
[\x \;,\; {\bar D}_i^\ad] \; \propto \; {\bar D}_j^\bd ~,
\label{4Dmaster0}
\ee   
and therefore
\be
{\bar D}_i^{\dt \a }\,\x^\b_j =0~, \qquad
{\bar D}_i^{\dt \a } \x^{\dt \b \b} = 4{\rm i} \, \ve^{\dt \a{}\dt \b} \,\x^\b_i~.
\label{4Dmaster}
\ee
The spinor covariant derivatives are assumed to obey the 
anti-commutation relations 
\bea
\{ D^i_\a, D^j_\b \} = \{ {\bar D}_{\dt \a i} , {\bar D}_{\dt \b j} \}=0~,
\qquad \{ D^i_\a, {\bar D}_{\dt  \b  j } \} =-2{\rm i} \,(\s^c)_{\a \dt \b}\, \pa_c~.
\label{spinor-c-d}
\eea

It follows from eqs. (\ref{4Dmaster0})  and (\ref{4Dmaster}) 
\be
[\x \;,\; D^i_\a ] = - (D^i_\a \x^\b_j) D^j_\b
= { \o}_\a{}^\b  D^i_\b - \frac{1}{\cN}
\Big( (\cN-2) {\s} + 2 \bar{ {\s}}  \Big) D^i_\a
- {\L}_j{}^i \; D^j_\a \;.
\label{4Dmaster2} 
\ee
Here the parameters of `local' Lorentz ${\o}$ 
and
scale--chiral ${\s}$ transformations 
are
\be
{\o}_{\a \b}(z) = -\frac{1}{\cN}\;D^i_{(\a} \x_{\b)i}\;,
\qquad {\s} (z) = \frac{1}{\cN (\cN - 4)}
\Big( \hf (\cN-2) D^i_\a \x^\a_i - 
{\bar D}^{\dt \a}_i {\bar \x}_{\dt \a}^{ i} \Big)
\label{lor,weyl}
\ee
and turn out to be chiral
\be
{\bar D}^{\dt \a}_{ i} {\o}_{\a \b}~=~ 0\;,
\qquad {\bar D}^{\dt \a}_{ i} {\s} ~=~0\;.
\ee
The parameters ${\L}_j{}^i$ defined by
\be
{\L}_j{}^i (z) = -\frac{\rm i}{32}\Big(
[D^i_\a\;,{\bar D}_{\dt \a j}] - \frac{1}{\cN}
\d_j{}^i  [D^k_\a\;,{\bar D}_{\dt \a k}] \Big)\x^{\dt \a \a}~, \qquad
{\L}^\dag = - {\L}~, \qquad  {\rm tr}\,{\L} = 0
\label{lambda}
\ee
correspond to `local' $SU(\cN )$ transformations.
One can readily check the identity 
\be
D^k_\a {\L}_j{}^i = -2 \Big( \d^k_j D^i_\a 
-\frac{1}{\cN} \d^i_j D^k_\a  \Big) {\s}~.
\label{an1}
\ee

The explicit expressions for the components  $\x^a(z) $ and $\x^\a_i(z)$ 
of an arbitrary  superconformal Killing vector can be found in \cite{K}, eq. (3.15).

\small{

}

\end{document}